\documentclass[aps,prd,10pt,twocolumn,nofootinbib,floatfix,
  noshowpacs,preprintnumbers]{revtex4-1}
\usepackage{bm}
\usepackage{epsfig}
\usepackage{amsmath}
\usepackage{color}
\usepackage[dvipsnames]{xcolor}
\usepackage[colorlinks=true,breaklinks=true]{hyperref}
\hypersetup{allcolors=[rgb]{0.0 0.0 0.6},linkcolor=[rgb]{0.75 0.05 0.05}}
\usepackage[normalem]{ulem}

\newcommand{\exclude}[1]{}

\begin{document}

\title{Neutrino Flavor Conversions in High-Density Astrophysical and
Cosmological Environments}

\author{Francesco Capozzi}
\affiliation{Instituto de F\'isica Corpuscular, Universidad de Valencia \& CSIC, Edificio Institutos de Investigaci\'on, Calle Catedr\'atico Jos\'e Beltr\'an 2, 46980 Paterna, Spain}
\author{Ninetta Saviano}
\affiliation{Scuola Superiore Meridionale, Universit\`a degli studi di Napoli “Federico II”, Largo San Marcellino 10, 80138 Napoli, Italy}
\affiliation{INFN - Sezione di Napoli, Complesso Univ. Monte S. Angelo, I-80126 Napoli, Italy}

\date{February 2, 2022}

\begin{abstract}

 \centering

Despite being a well understood phenomenon in the context of current terrestrial experiments, neutrino flavor conversions in dense astrophysical environments probably represent one of the most challenging open problems in neutrino physics. Apart from being theoretically interesting, such a problem has several phenomenological implications in cosmology and in astrophysics, including the primordial nucleosynthesis of light elements abundance and other cosmological observables,  nucleosynthesis of heavy nuclei and the explosion of massive stars. In this review, we briefly summarize the state of the art on this topic, focusing on three environments: early universe, core-collapse supernovae and compact binary mergers.

\end{abstract}

\maketitle

\section{Introduction}
\label{sec:Introduction}

Neutrino flavor conversions, or oscillations, are a genuine quantum mechanical phenomenon, for which a flavor eigenstate $\nu_\alpha$  ($\alpha=e,\mu,\tau$) is converted to $\nu_\beta$  ($\beta\ne\alpha$) during propagation, due to it being an admixture of different mass (or propagation) eigenstates $\nu_i$ ($i=1,2,3$). This has been firmly established experimentally with atmospheric \cite{RevModPhys.88.030501} and solar neutrinos \cite{RevModPhys.88.030502}, proving that neutrinos are massive particles and leading to the award of the Nobel Prize in Physics 2015 to the experiments involved in this discovery. Among the parameters describing such a phenomenon, the three mixing angles $(\theta_{12},\theta_{13},\theta_{23})$ and the two mass differences $(\Delta m^2_{21},\Delta m^2_{31})$ have been measured with a precision between 1 and 10\% \cite{Capozzi:2021fjo,Gonzalez-Garcia:2021dve,deSalas:2017kay}. The mass ordering (normal for $\Delta m^2_{31}>0$ or inverted for $\Delta m^2_{31}<0$) is still unknown, but there is a $3\sigma$ hint in favor of the normal one. The CP phase $\delta$, which leads to the violation of the CP symmetry if $\delta\ne 0,\pi$, is also largely unknown, but the CP conserving values seem to be disfavoured at 90\% confidence level and the best fit seems to lie in the range [$\pi,2\pi$].  Despite the presence of some partially unknown parameter and of few anomalies \cite{LSND:2001aii,MiniBooNE:2018esg,Abdurashitov:2005tb,Kaether:2010ag,Mention:2011rk}, which still need to be confirmed and require further investigations, flavor conversions relevant in current experiments are theoretically under control.

Our understanding becomes less solid the higher the density of neutrinos being considered. In particular, when the interactions among neutrinos are no longer negligible, their flavor evolution becomes deeply non linear and cannot be treated in the standard way used in the context of terrestrial experiments. The environments where such conditions occur are the early universe, core-collapse supernovae and merger events between two compact astrophysical objects. Apart from representing an extremely interesting theoretical problem to solve, flavor conversions in dense environments have a much deeper relevance. In particular, neutrino flavor conversions in the  early universe are a fascinating problem involving collisional damping, refractive effects from charged leptons, and neutrino self-interactions \cite{Dolgov:2002ab}. In this context, there is  a strong interest for  active-sterile neutrino oscillation  in the early universe in a broad range of mass for the sterile neutrino. Concerning supernovae, how neutrinos are distributed among all flavors affects the amount of energy deposited by these particles just below the shock wave in a supernova, potentially changing a successful explosion into a failed one, or viceversa. Furthermore, the large amount of information carried by supernova neutrinos can be fully exploited by their detection only if we can precisely predict both the original fluxes for each flavor and how they are modified during propagation.

In this review our goal is to provide a short summary of the state of the art of the available literature on flavor conversions in dense environments, focusing on both the theoretical understanding of the equations of motion and applications to phenomenological aspects of the environments under consideration. In Section \ref{sec:eq_motions} we present a common theoretical framework based on a Boltzmann equation of the neutrino density matrix. In Section \ref{sec:Early_Universe}, \ref{sec:Supernovae} and \ref{sec:Compact_binaries} we discuss old and recent results concerning flavor conversions in the early universe, core-collapse supernovae and compact binary mergers, respectively. Finally we give our conclusions and future perspectives.

\section{Equations of motions}
\label{sec:eq_motions}

The simple formalism of the Schr\"odinger equation for the treatment of neutrino mixing is not very practical whenever we are concerned with the evolution of a statistical ensemble of neutrinos 
simultaneously mixing and scattering in a medium.  In such cases the evolution cannot be easily understood 
in terms of the propagation of a beam. Indeed,  while this treatment allows to obtain the 
transition probabilities between single-particle states and to study the oscillations between
field amplitudes,it cannot be directly applied to many-particle states. 
Specifically, we need to handle the interactions between the medium and the coherent
superposition of states that are involved. In presence of a propagation medium it is important
to follow  the evolution of the whole ensemble, including those particles scattered out of the beam. Two types of effects have to be considered. One concerns  the  refractive effect namely the potential due to coherent forward scattering on charged leptons present in ordinary matter \cite{Wolfenstein:1977ue,Mikheyev:1985zog}. The second effect is due to collisions which destroy the coherence of the evolution and  can influence the  behavior of the mixing process \cite{Stodolsky:1986dx}.

In addition to the interactions with the external medium,  one has also to consider the  interactions of neutrinos among themselves (\emph{self-interaction}). Indeed, as pointed out by Pantaleone in the 1992 \cite{Pantaleone:1994ns} , in the deepest region of the Supernova (SN) and in the early universe, the neutrinos gas is so dense that the neutrinos themselves form a background medium leading to intriguing non-linear effects in the neutrino flavor conversions.

A proper treatment of all  these effects requires to exploit the density matrix formalism, in which  mixed quantum states for neutrinos and possible loss of
coherence due to real collisions are described  together. Strictly following the derivation of  \cite{Sigl:1993ctk} and also \cite{Raffelt:1991ck, Raffelt:1992uj,Dolgov:1980cq, Barbieri:1990vx}, an
ensemble of neutrinos and antineutrinos can be characterized by the $n \times n$  density
matrices $\varrho_{\mathbf{p,x}}(t)$ (and an analogue for antineutrinos $\overline{\varrho}_{\mathbf{p,x}}(t)$) defined  as 
\begin{align}
 \small{
\varrho_{\mathbf{p,x}}=
\left(\begin{array}{ccc} 
\varrho_{ee} &  \varrho_{e \mu} & \varrho_{e\tau} \\
\varrho_{\mu e}  & \varrho_{\mu \mu} &  \varrho_{\mu \tau} \\
\varrho_{\tau e} &\varrho_{\tau \mu} &\varrho_{\tau \tau}
\end{array}\right)}~, 
\label{eq:rho}
\end{align}
 where the diagonal elements of $\varrho_{\mathbf{p,x}}$  are the usual particle distribution functions (occupation numbers) for the corresponding neutrino species, while the off-diagonal ones encode phase information and vanish for zero mixing. In general $\varrho_{\mathbf{p,x}}$ depends on the momentum ${\bf p}$ and on the coordinate 
${\bf x}$).

The entries of $\varrho_{\mathbf{p},\mathbf{x}}$ in Eq. \ref{eq:rho} are intended as the expectation values of the neutrino field bilinears. Under this approximation, labelled as the mean field approach, one neglects possible entanglement effects. Despite the wide use of this approach, it has been debated in a series of papers \cite{Friedland:2003eh,Friedland:2003dv,Balantekin:2006tg,Pehlivan:2011hp,Birol:2018qhx,Patwardhan:2019zta,Cervia:2019res,Rrapaj:2019pxz,Roggero:2021asb,Roggero:2021fyo}.

In this review we will focus on the results obtained in the mean field approximation. In this case the flavor evolution 
for the density matrices $\varrho_{\bf p,x}$ and $\bar\varrho_{\bf p,x}$   in dense environments is governed by the Boltzmann collision 
equations 
\begin{equation} 
 \hat{L}[\varrho_{\bf p,x} ]= 
 -i[{\Omega}_{\bf p,x},\varrho_{\bf p,x} ] 
 +\hat{\mathcal{C}}[\varrho_{\bf p,x}]
\label{eq:boltz} 
\end{equation} 
where at r.h.s.  ${\Omega}_{\bf p,x}={\Omega}^{\rm vac}_{\bf p} + {\Omega}^{\rm ref}_{\bf p,x} $ and ${\Omega}^{\rm ref}_{\bf p,x}={\Omega}^{\rm mat}_{\bf p,x}+{\Omega}^{\rm \nu\nu}_{\bf p,x}$. ${\Omega}^{\rm mat}_{\bf p,x}$ describes the interactions with all other particles of the medium except for neutrinos and ${\Omega}^{\rm \nu\nu}_{\bf p,x}$ represents the ``self-interactions" with the other neutrinos in the medium and, depending on the neutrino ensemble $\varrho_\mathbf{p,x}$ itself,
it makes the problem non-linear. Finally, the last term at the right-hand side (r.h.s) of the Eq. \ref{eq:boltz} is the second order in the perturbative expansion ($\propto G^2_F$), known as collisional
term, responsible for the breaking of the coherence of the neutrino ensemble.
The l.h.s. of Eq. \ref{eq:boltz} contains the  Liouville operator
\begin{equation}
 \hat{L}[\varrho_{\bf p, \bf x}] = \partial_t\varrho_{\bf p,x} 
 +{\bf v}_{\bf p}\cdot{\nabla}_{\bf x}\,\varrho_{\bf p,x} 
 +\dot{\bf p}\cdot{\nabla}_{\bf p}\,\varrho_{\bf p,x},  
\end{equation} 
which includes temporal  evolution and spatial transport phenomena. 
In particular, the first term represents an explicit time dependence, the second a drift caused by the 
particles free-streaming, and the third the effects of external macroscopic forces, for example gravitational deflection. 
 The equation for $\bar\varrho_{\bf p,x}$ is similar to the equation \ref{eq:boltz},  where inside the commutator the relative sign of ${\Omega}^{\rm vac}$ and ${\Omega}^{\rm ref}$ changes.

\section{Flavor Conversions in the early universe}
\label{sec:Early_Universe}

The evolution in time in the form of  Boltzmann-like equations  applied to the early universe, safely considered  isotropic and homogeneous at large scale,  reduces to 
\begin{equation}
 \partial_t\varrho_{\bf p} = -i[{\Omega}_{\bf p},\varrho_{\bf p}]
 +\hat{\mathcal{C}}[\varrho_{\bf p}]~,
 \end{equation}
where
\begin{equation}
\partial_{t} \rightarrow  \partial_{t} - Hp~\partial_{\mathbf{p}}
\end{equation}
with $H$ the Hubble parameter which encodes the information about the universe expansion.
The quantity  $\Omega_p$  includes the \emph{vacuum term}, the \emph{Refractive matter term} and the \emph{neutrinos self-interactions}.

For the Refractive matter term,  since the electron-positron density is expected  to be of the same order of the baryons-antibaryons ones which is subdominant. Therefore, it is necessary to consider the higher order  which depends on the sum of the electron and positron energy densities.
Concerning the self-interaction term,  the contributions to neutrino-neutrino forward scatterings come at leading order from a term  $\Omega_{\mathbf{p,x}}^{\nu \nu(0)}\propto G_F (\varrho - \overline{\varrho})$, and at higher order from a term $\Omega_{\mathbf{p,x}}^{\nu \nu(1)}\propto \frac{G_F}{m^2_Z}(\varrho + \overline{\varrho})$ \footnote{In order to include the correction due to the non-local nature of the Z boson propagator which mediate forward scattering on neutrinos of the same species.}.  In absence of a neutrino-antineutrino asymmetry, as expected in the standard case, the only contribution is given by $\Omega_{\mathbf{p,x}}^{\nu \nu(1)}$ which is subleading for the neutrino evolution, since its numerical value is small. Conversely, in the case of large neutrino asymmetries, $\Omega_{\mathbf{p,x}}^{\nu \nu(0)}$  becomes important and the evolution is  dominated by the effect of \emph{synchronized} oscillations, i.e. the self-potential forces all neutrino modes to follow the same oscillation pattern \cite{Dolgov:2002ab}.
Finally, the \emph{collisional term}  $\hat{\mathcal{C}}[\varrho_{\bf p}]~$ at high temperature of the primordial plasma,  is very important since it  breaks the coherence of the neutrino ensemble. It damps the off diagonal terms of the density matrix $\varrho_{\mathbf{p}}$ and it pushes the diagonal terms towards  their equilibrium distributions. 

In the case of instantaneous neutrino decoupling (at temperature around 1 MeV) from the primordial plasma of particles, the relic neutrinos of each flavor have the same momentum distributions making the effect of the oscillations irrelevant  except for an exchange of  equal  neutrino spectra.
However, the neutrino oscillations become important in some situations in which unequal neutrino distributions arise, such as small flavour-dependent distortions due to not-instantaneous decoupling, non zero neutrino-antineutrino asymmetry, or sterile species mixing with the active ones \cite{Lesgourgues:2013sjj,Dolgov:2002wy}. In such situations, neutrino oscillation could modify the non-electromagnetic  contribution of the neutrino heating to the total relativistic energy density, defined in term of the effective number of neutrinos species

\begin{equation}
\epsilon_R= \epsilon_{\gamma} \left( 1 + \frac{7}{8} \left(\frac{4}{11} \right) ^{4/3}~ N_{\textrm{eff}} \right)\label{Neff}
\end{equation}

where $\epsilon_R$ and $\epsilon_{\gamma}$ are the total energy density of radiation and the energy density of photon, respectively. The factor 4/11 comes form  the heating of  the photons due to the $e^+ - e^-$ annihilation.
 In the case of small non-thermal distortions  plus oscillations, the value of this parameter is estimated to be $\rm N_{eff}= 3.046$ \cite{Mangano:2005cc}. The slightly excess  with respect to the value $3$ is to due  to the non instantaneous neutrino  decoupling  thanks to which  neutrinos share a small part of the entropy release after the $e^+ - e^-$  annihilation. More recently new calculations,   where finite-temperature effects in the quantum electrodynamics plasma and a full evaluation of the neutrino–neutrino collision integral are taken into account, provide a number very close to the original one \cite{Bennett:2020zkv,Froustey:2020mcq}.

An important implication of active neutrino oscillations at temperature of  MeV of the Universe is the evolution of a possible neutrino-antineutrino asymmetry, denoted by 
\begin{equation}
L_{\nu_\alpha}= \frac{n_{\nu_\alpha}-n_{\bar{\nu_\alpha}}}{n_\gamma}.
\end{equation}
Considering the very small value of the  baryonic asymmetry, $\eta_\beta = (n_B-n_{\bar B})/n_{\gamma} \simeq 6 \times 10^{-10}$, it is reasonable to expect, for  the charge neutrality, the same order of magnitude for the charged lepton asymmetry. In the neutrino case, being neutral  particles, the constrains on $L_{\nu_\alpha}$ are quite loose, allowing values for $L_{\nu_\alpha}$ order of magnitude larger then $\eta_\beta$. The presence of  $L_{\nu_\alpha}$ implies a degeneracy parameter in the neutrinos spectra, $\xi_{\alpha}=\mu_{\nu_\alpha}/T_\nu$. A significant value of a cosmological  neutrino-antineutrino asymmetry implies an  extra contribution for $\rm N_{eff}$ without  to introduce additional relativistic degrees of freedom. An important issue is to compute the evolution of $L_{\nu_\alpha}$ in the epoch of the Universe before Big Bang Nucleosynthesis (BBN). A combined analysis of active neutrino flavor oscillations and BBN has led
to an almost standard value for the effective number also in the presence of neutrino asymmetries, with  $N_{\rm eff}< 3.2$ at $95 \%$ C.L. \cite{Mangano:2011ip}.

\subsection{Active-sterile neutrino oscillations}

The existence of sterile neutrinos is investigated in a very broad range of mass, from the GUT (Grand Unification Theory) scale to the eV scale. Some of them  are theoretically very well motivated, other are  more suggested by possible experimental hints. In this context,  a special and interesting case is represented by neutrino oscillations among active and sterile neutrinos at different mass scales.

\textbf{eV sterile neutrinos}-Light sterile neutrinos with a mass around $\sim 1$ eV, which are mixing with the active ones, have been suggested to solve different anomalous results observed in $\nu_\mu\to\nu_e$ (LSND \cite{LSND:2001aii} and MiniBooNE \cite{MiniBooNE:2018esg}) and $\nu_e\to\nu_e$ (SAGE \cite{Abdurashitov:2005tb} and GALLEX \cite{Abdurashitov:2005tb}) short-baseline oscillations, as well as in $\bar{\nu}_e\to\bar{\nu}_e$ reactor neutrino experiments \cite{Mention:2011rk}.
Many analyses have been performed  to explain the anomalies and
scenarios with one (dubbed ``3+1'') or two (``3+2'') sub-eV sterile neutrinos~\cite{Giunti:2011gz,Kopp:2013vaa,Dentler:2018sju, Gariazzo:2017fdh,Diaz:2019fwt,Boser:2019rta} have been proposed to fit the different data. 
The search for sterile neutrinos in laboratory experiments is still undergoing. The $\nu_\mu\to\nu_e$ has been recently tested in MicroBooNE \cite{MicroBooNE:2021nxr,MicroBooNE:2021rmx,MicroBooNE:2021sne,MicroBooNE:2021jwr}, whose current data cannot exclude the full parameter space hinted at by MiniBooNE \cite{Arguelles:2021meu} and if combined with MiniBooNE leads to a preference for a 3+1 scenario with a best fit at $\Delta m^2\sim 0.2$ eV$^2$ and very large mixing \cite{MiniBooNE:2022emn}. The $\nu_e\to\nu_e$ has now been updated by the data collected by BEST \cite{Barinov:2021asz}, which strongly favors rather  large  mixing  angle  and  masses larger than 1 eV$^2$. Concerning the reactor anomalies, a recent reevaluation of the neutrino fluxes \cite{Kopeikin:2021ugh} has led to new global analyses \cite{Giunti:2021kab,Berryman:2021yan} showing that combined reactor data are consistent with no sterile-neutrino oscillations. Despite the recent progress in all oscillation channels, new experimental data is still needed to fully assess the sterile neutrino option.

\begin{figure}
  \centering
  \includegraphics[width=0.40\textwidth]{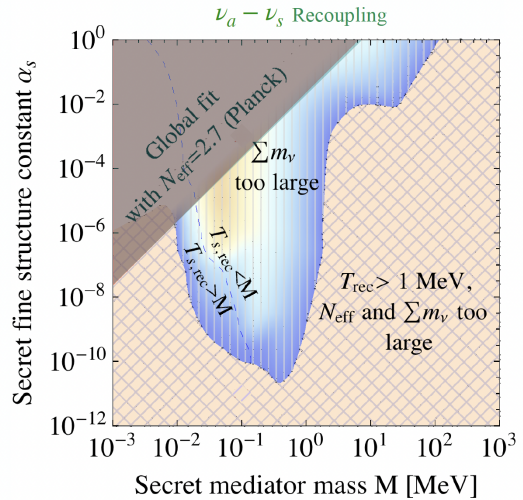}
  \caption{The parameter space of the secret interactions model mediated by a vector  boson of mass M and the corresponding fine structure constant $\alpha_s$ ($\alpha_s=g^2_s/(4 \pi)$). A sterile neutrino mass $m_s$ = 1 eV and a vacuum mixing angle $\theta_s$= 0.1 are assumed. The colored regions are ruled out by different cosmological observations. This figure is adapted from \cite{Chu:2018gxk}.}
  \label{fig:exclusionplot}
\end{figure}

In this context, cosmological observations represent a valid complementary tool to  probe these scenarios, being sensitive to  the number of neutrinos and to their mass 
at eV scale \cite{Hagstotz:2020ukm}. In fact, the sterile  states  can be produced in the early universe via oscillations  with active neutrinos and  can modify cosmological observables. In particular, adding extra contribution to the radiation content in the Universe,  expressed in terms of the effective number of  neutrino species, $N_{\rm eff}$, has indeed a strong impact on both  BBN primordial yields and the Cosmic Microwave Background (CMB) anisotropy. Moreover, the mass of sterile species can impact  the  structure formation (LSS) spectrum \cite{Mangano:2011ar,Giusarma:2011ex,Lesgourgues:2012uu, 
Riemer-Sorensen:2013iql,Hagstotz:2020ukm}. From recent 
astrophysical data of Deuterium and Helium, as well as last 
constraints on radiation and mass, even a single extra 
thermalized sterile neutrino with  mass  $m\sim 1$ eV appears to be inconsistent BBN,  CMB and LSS data~\cite{Cooke:2017cwo,Grohs:2019cae,Hsyu:2020uqb, Fields:2019pfx,Planck:2018vyg,Mirizzi:2013kva,Gariazzo:2019gyi}. A possible escape-route to reconcile the eV sterile neutrino interpretation with cosmology would be to suppress the sterile neutrino thermalization and therefore  non standard scenarios have to be invoked to alleviate the tension.   One of the proposed mechanism  is to consider a primordial asymmetry $L_{\nu}$ between active neutrinos and antineutrinos which implies an additional “matter term potential” in the equations of motion. If sufficiently large, it can block the active-sterile flavor conversions via the in-medium suppression of the mixing angle. However, this term can also generate MSW-like resonant flavor conversions among active and sterile neutrinos, enhancing the sterile neutrino production. A detailed study of the kinetic equations for  active-sterile neutrino oscillations is therefore mandatory to assess which of the two effects dominates. Recent studies have shown that asymmetry $L_{\nu}\sim 10^{-2}$ is required in order to achieve a sufficient suppression of the sterile neutrino abundance, \cite{Mirizzi:2012we}. However, such a large value delays the oscillations at temperature close to neutrino decoupling one inducing distortions on the active neutrino spectra with   non-trivial effects on BBN. Consequently, the tensions with cosmology re-emerges \cite{Saviano:2013ktj}. Recently, different groups  have proposed  and investigated  an alternative method to suppress the sterile neutrino production  based on the introduction of new ``secret'' self- interactions among sterile neutrinos, mediated by new gauges boson \cite{Chu:2015ipa,Hannestad:2013ana}. Also in this case , the self-interactions would generate a matter potential in the flavor evolution equations potentially  suppressing the sterile neutrino abundance. Depending on the selected theoretical model, we have more or less severe constraints on light sterile neutrinos induced by the cosmological data.  BBN, CMB and neutrino mass bounds strongly constrain the model with vector boson, disfavoring all the values of the mass of the new gauge boson. Indeed, while the effective temperature-dependent potential produced by secret interactions can efficiently suppress active–sterile neutrino mixing in the early universe down to lower temperatures, the momentum spectra of active neutrinos will result  distorted due to delayed oscillations with impact on the production of BBN yields \cite{Saviano:2014esa}.
Moreover, efficient collisional production of $\nu_s$ can still  occur  (depending on the neutrino temperature and on the mass of the mediator)   violating the  cosmological constraints on CMB and neutrino mass bound of cosmological structures \cite{Mirizzi:2014ama,Forastieri:2017oma} . In particular, scattering processes mediated by the new particles can be strongly enhanced by the s-channel resonance  and by collinear enhancement in the forward direction.  For the mass and mixing parameter preferred by laboratory anomalies, all values of the vector boson and the corresponding coupling constants are disfavored \cite{Chu:2018gxk}, as shown in Fig. \ref{fig:exclusionplot}.

The situation can be in part different in the case of a very light (or even massless) pseudoscalar particle mediating the secret interactions, due to its late time phenomenology, in particular since the collisional term  becomes relevant at late times \cite{Archidiacono:2014nda,Archidiacono:2016kkh,Archidiacono:2020yey}.
 Performing a series of extensive analyses fitting various combinations of CMB  data and combining with other cosmological data, it is still debated if eV sterile neutrinos could be accommodated within the pseudoscalar interaction model \cite{Corona:2021qxl}.
 Another chance to possibly alleviate the eV-sterile neutrino controversy could be to consider minimal dark energy models which  modify only late time physics\cite{DiValentino:2021rjj}.\\

 \textbf{keV sterile neutrinos}- KeV sterile neutrinos could represent a valid candidate of Dark Matter  since they are neutral,  cosmologically stable, sufficiently non-interacting,  and massive \cite{Drewes:2016upu,Abazajian:2017tcc,Boyarsky:2018tvu}. A renewed interest has been recently sparked by new appealing and  controversial indications for an unidentified X-ray line  at 3.5 keV in the spectra of a large number of galaxy clusters and in Andromeda, possibly consistent with a radiative decay of 7 keV sterile neutrinos. \cite{Bulbul:2014sua,Boyarsky:2014jta}.
Sterile neutrinos in this mass range are usually classified as warm candidates of dark matter because, when they become non-relativistic,
their free-streaming suppresses the matter density perturbations at the dwarf galaxies scales, more in
agreement with the observations and cosmological simulations. However, the free-streaming length not only
depends on the particle mass, but also on the momentum, and then on the particle phase-space distribution \cite{Murgia:2017lwo}.
Depending on the production mechanism for sterile neutrinos, the distribution function can be more or less
warm, with a different impact on the structure formation. The most natural production mechanism to produce  keV sterile neutrinos is  by mixing with active species. The 
mixing angle which  allows the decay of  sterile neutrinos  also enables an efficiently production of sterile species in the early universe.
 The \textit{non-resonant production},  linked to DM by Dodelson and Widrow \cite{Dodelson:1993je}, is a collision-dominated production via neutrino oscillations,
which could gradually produce enough sterile neutrinos to explain the observed amount of DM, in according to the production rate
\begin{equation}
\Gamma_{\rm prod}= \frac{1}{2}\sin^2 2\theta_m \Gamma
\end{equation}
driven by the active neutrino interaction
rate 
\begin{equation}
\Gamma \sim G_F^2 T^5. \label{gamma}
\end{equation}

However, in the absence of new physics,  this mechanism   results to be  similar to the hot DM case and  in tension with various current astrophysical observations \cite{Tremaine:1979we,Boyarsky:2008ju,Gorbunov:2008ka,Horiuchi:2013noa,Ng:2019gch}.  However, this  situation could change in the case of a \textit{resonant enhancement}  in the active-sterile transitions due to  a lepton number asymmetry $L_{\nu}$ possibly present in the early universe. Indeed, as pointed out by Shi-Fuller \cite{Shi:1998km},  a suitable value of  such an asymmetry  produces  sterile neutrinos at a  specific combination of momentum and temperature resulting in colder DM distribution and so improving the compatibility with astrophysical and cosmological data. Anyway, also this scenario is  strongly constrained \cite{Cherry:2017dwu,Wang:2017hof,Schneider:2017qdf,Schneider:2016uqi,Drewes:2016upu} and the 7 keV sterile neutrino interpretation of the observed 3.5 keV line is not favoured \cite{Dessert:2018qih,Enzi:2020ieg}.
 As recently speculated by different groups, the allowed mass and mixing parameter space could be extended in presence of new interactions among only active neutrinos \cite{DeGouvea:2019wpf,Kelly:2020pcy} or among only sterile species \cite{Johns:2019cwc} which imply extra terms in the interaction rate $\Gamma$ (eq. (\ref{gamma})).\\
 
 \textbf{MeV sterile neutrinos}- They emerge rather naturally in theories beyond the Standard Model, like low scale seesaw models in the Neutrino Minimal Standard Model being related to fundamental open  problems of particle physics such as the origin of neutrino mass, the baryon asymmetry in the early universe and the nature of dark matter\cite{Asaka:2005an,Asaka:2005pn}. Depending on the mixing with active neutrinos, the parameter space of sterile neutrino in MeV mass range is  strong constrained by: collider and beam-dump experiments \cite{Alekhin:2015byh},  searches of decays of D mesons and $\tau$ leptons \cite{Chun:2019nwi,Orloff:2002de} and  core-collapse supernova limits due to energy loss  which would short the observed neutrino burst \cite{Dolgov:2000jw,Fuller:2008erj,Mastrototaro:2019vug}. Further and complementary constraints can also come from cosmological observations. Indeed, sterile neutrinos produced in the early universe via collisional processes involving active neutrinos, can decay into lighter species injected into the primordial plasma altering both  the effective number of neutrino species $N_{\rm eff}$ and  
 the abundance of the primordial yield. Solving the exact Boltzmann equation for sterile and active neutrino evolution  is possible to set  constraints on the mixing angles or lifetimes \cite{Mastrototaro:2021wzl}.

\section{Flavor Conversions in Core-Collapse Supernovae}

\label{sec:Supernovae}
 
Core-collapse supernovae are the  final  explosion  of  a  star  with  a  mass greater than 8 $M_\odot$. After extinguishing all the nuclear fuel, the core of the star begins to collapse under the effect of gravity until it reaches the Chandrasekhar limit. Then the  density rapidly  increases  until  the  core  reaches  the  nuclear density  ($3\times 10^{14}$ g/cm$^3$), when matter becomes incompressible  and  the  collapse  stops,  producing  a  rebound \cite{Colgate:1960zz}. The bounce generates a shock wave travelling outwards and ejecting outer layers. This shock may stall without driving off the stellar mantle and envelope.  According to the ``delayed explosion scenario" \cite{Colgate:1966ax,Bethe:1985sox}, since 99\% of the liberated gravitational energy ($10^{53}$ ergs) is in the form of neutrinos \cite{Gamow:1940eny,Gamow:1941gis} such particles might deposit enough energy below the shock wave in order to revitalize the explosion. 

Neutrino emission can be divided in three main phases (see e.g., \cite{Mirizzi:2015eza,Janka:2017vlw,Burrows:2020qrp} for detailed overviews). The neutronization burst lasts $\sim10$ ms  and  consists in  the  rapid electron capture on dissociated nuclei.  This leads to a sudden rise in the luminosity of $\nu_e$ up to $10^{53}$ erg/s, while for the other flavors it is negligible. The accretion phase lasts a few hundreds of ms and during this time material continues to fall onto the core and accretes on it, leading to a flux of neutrinos and antineutrinos of all species. The high density traps neutrinos \cite{Mazurek1,Freedman:1973yd,Mazurek2,Mazurek3}, which escape from the last scattering surface, which is usually called the neutrinosphere. In the cooling phase the remaining proto-neutron star cools by neutrino emission of all flavors. During this phase the luminosity is approximately equal for all flavors.

Apart from inside the neutrinosphere, in general the propagation of supernova neutrinos is not isotropic. Consequently, the self-interaction potential has the following form
\begin{equation}
    \Omega_{\mathbf{p,x}}^{\nu\nu}=\sqrt2 G_F
  \int\,\frac{\mathrm{d}^3\mathbf{q}}{(2\pi)^3}\,(\varrho_{\mathbf{q,x}}-\bar{\varrho}_{\mathbf{q,x}})  ( 1 - \mathbf{q}\cdot\mathbf{p})\,,
  \label{eq:self_potential_SN}
\end{equation}
whereas for the isotropic case of the early universe the $(1-\mathbf{p}\cdot\mathbf{q})$ term is absent. The phenomenology of flavor conversions is strongly dependent on the relative size of $\Omega_{\mathbf{p,x}}^{\nu\nu}$ and $\Omega_{\mathbf{p,x}}^{\rm mat}$, which changes as a function of both time after bounce and radial distance from the centre. In general, a few thousands kilometers from the centre of the supernova the neutrino density is low enough to make $\Omega_{\mathbf{p,x}}^{\nu\nu}$ negligible. In this case $\Omega_{\mathbf{p,x}}^{\rm mat}$ can give rise to the MSW resonance \cite{Wolfenstein:1977ue,Mikheyev:1985zog}. Deeper inside the supernova, $\Omega_{\mathbf{p,x}}^{\nu\nu}$ can be of the same order or even larger than the matter potential, making flavor conversions a deeply non linear phenomenon and giving rise to the so called collective effects. Such a name stems from the fact that neutrinos can change their flavor in a coherent fashion, regardless of their original energy. 

A remark is in order. The role of flavor conversions in a SN is not important just for a correct determination of the expected neutrino signal, but also in the context of astrophysical processes, such as the nucleosynthesis of heavy elements \cite{Burbidge:1957vc,Qian1993,Chakraborty:2009ej,Duan:2010af,Xiong:2020ntn} and the explosion itself \cite{Fuller1992}, since neutrino energy deposition is flavor dependent. Below we briefly review all types of flavor conversions happening in a supernova, focusing on both theoretical and phenomenological aspects.

\subsection{MSW resonances}

An MSW resonance occurs when the vacuum oscillation frequency is equal to the spatially dependent matter potential  \cite{Wolfenstein:1977ue,Mikheyev:1985zog}. Such a condition is equivalent to $\omega_E=\pm V_{\rm mat}(\mathbf{x})$, where $\omega_E=\Delta m^2/(2E_\nu)$ is the vacuum oscillation frequency of neutrinos, and $V_{\rm mat}(\mathbf{x})=\sqrt{2}G_FN_e(\mathbf{x})$ is the potential from coherent forward scattering on electrons in a medium, with $N_e(\mathbf{x})$ being the spatially dependent electron number density. The $+$ ($-$) sign refers to neutrinos (antineutrinos). If $\Delta m^2=\Delta m^2_{31}$, then the resonance occurs for neutrinos in normal mass ordering and for antineutrinos in inverted mass ordering. If $\Delta m^2=\Delta m^2_{21}>0$ then the resonance can only happen for neutrinos. The final survival probability of electron neutrinos is given by \cite{Dighe:1999bi}
\begin{equation}
P_{ee}(E_\nu) = \left\lbrace
                \begin{array}{llll}
                  \sin^2\theta_{12}P_H\quad (\nu,\text{ NO})\\
                   \sin^2\theta_{12}\quad (\nu,\text{ IO})\\
                   \cos^2\theta_{12}\quad (\bar{\nu},\text{ NO})\\
                   \cos^2\theta_{12}P_H\quad (\bar{\nu},\text{ IO})\\
                \end{array}
              \right.
\label{MSW_Pee}
\end{equation}
where $P_H$ represents the probability that a given mass eigenstate is transformed to a different one when crossing the resonance point. This is also known as the crossing probability and it is equal to 0 when the propagation is adiabatic, whereas it must be determined (numerically or analytically) in the opposite case. In particular, when the resonance crosses the shock wave, which introduces a sharp variation in the electron density, the propagation becomes non adiabatic. This is a time dependent feature, whose potential observation has been studied in \cite{Schirato:2002tg,Fogli:2003dw,Fogli:2004ff,Tomas:2004gr,Dasgupta:2005wn,Choubey:2006aq,Kneller:2007kg,Friedland:2020ecy}.

The presence of turbulence in the matter density can introduce stochastic fluctuations and thus induce random  temporal variations in the occurrence of the MSW resonance \cite{Fogli:2006xy,Friedland:2006ta,Kneller:2010sc,Lund:2013uta,Loreti:1995ae,Choubey:2007ga,Benatti:2004hn,Kneller:2013ska}. If the fluctuations are of sufficiently large amplitude they might lead to $P_H \to 1/2$ \cite{Fogli:2006xy}.

\subsection{Collective effects}

In order to study the phenomenology of supernova neutrino flavor conversions in the presence of a large self-interaction potential one needs to solve Eq. \ref{eq:boltz}, with $\Omega_{\mathbf{p,x}}^{\nu\nu}$ given by Eq. \ref{eq:self_potential_SN}. Being a 7-dimensional system of coupled differential equations, it has been solved only imposing some simplifying symmetries.  

Even without solving the system of partial differential equations, some information can be obtained through a linear stability analysis \cite{Sawyer:2008zs,Banerjee:2011fj}. In this case one expands the density matrix as
\begin{equation}
\varrho_{\mathbf{p,x}}(t)=\frac{f_{\nu_e}+f_{\nu_x}}{2}+\frac{f_{\nu_e}-f_{\nu_x}}{2}\,
\begin{pmatrix}s_{\mathbf{p}}(t,\mathbf{x})&S_{\mathbf{p}}(t,\mathbf{x})\\S^*_{\mathbf{p}}(t,\mathbf{x})&-s_{\mathbf{p}}(t,\mathbf{x})\end{pmatrix}\,,
\end{equation}
where $f_{\nu_\alpha}$ represents the number of neutrinos with flavor $\alpha$.
Since neutrinos are produced in flavor eigenstates, one expects that initially $s_{\mathbf{p}}\ll S_{\mathbf{p}}$, which justifies an expansion of Eq. \ref{eq:boltz}  at linear order in $S_{\mathbf{p}}$. Then, one looks for plane wave solutions of the form $S_{\mathbf{p}}(t,\mathbf{x})=Q_\mathbf{p,k}\,e^{-i(k_0 t-\mathbf{k}\cdot\mathbf{x})}$, where $k_0$ is the temporal frequency of the flavor wave and $\mathbf{k}$ is its spatial wave vector. Plugging the plane wave expansions in Eq. \ref{eq:boltz} one arrives at a dispersion relation $\rm{det}[\Pi^{\mu\nu}]=0$, i.e. the determinant of a $4\times 4 $ matrix $\Pi^{\mu\nu}$ is equal to 0, where
\begin{equation}\label{eq:Pi}
\Pi^{\mu\nu}=
\eta^{\mu\nu}+\int_{0}^{+\infty}\frac{E^2dE}{2\pi^2}\int \frac{d\mathbf{v}}{4\pi}\,g_{E,\mathbf{v}}\,
\frac{v^\mu v^\nu}{v_\alpha\,k^\alpha-\omega_E}\,.
\end{equation}
Here, $v^\mu=(1,\mathbf{v})$, $k^\mu=(k^0,\mathbf{k})$,  $\mathbf{v}=\mathbf{p}/|\mathbf{p}|$ and $g_{E,\mathbf{v}}$ is defined as
\begin{equation}
g_{E,\mathbf{v}}=\sqrt{2}G_F[f_{\nu_e,E,\mathbf{v}}-f_{\nu_x,E,\mathbf{v}} - (f_{\bar{\nu}_e,E,\mathbf{v}}-f_{\bar{\nu}_x,E,\mathbf{v}}, )]\,,
\label{eq:gp}
\end{equation}
where $f_{\nu_\alpha,E,\mathbf{v}}$ is the energy and angular distribution of $\nu_\alpha$.
If for some real $k_0$ the dispersion relation is solved with an imaginary $\mathbf{k}$, the system has a spatial instability. Conversely, if for some real $\mathbf{k}$ the dispersion relation is solved with an imaginary $k_0$, the system has a temporal instability. This method allows one to know whether a system is unstable or not, possibly leading to significant flavor conversions. However, the final amount of flavor conversions to be expected is not accessible through this method.

A widely used classification of instabilities is done according to the growth rate of unstable modes, i.e. the imaginary part of $k$, ${\rm Im}(k)$. We define slow instabilities those having ${\rm Im}(k)\propto \sqrt{\omega_E \mu}$, while we define fast instabilities those having ${\rm Im}(k) \propto \mu$. The latter can reach ${\rm Im}(k)\sim O(1)$ cm$^{-1}$, if they occur deep inside the supernova core, where $\mu$ can be $O(10^5)$ cm$^{-1}$. This is why fast instabilities are usually considered to be the most likely to affect astrophysical processes occurring in a supernova, provided that they are triggered in the first place.

It has been proven in \cite{Morinaga:2021vmc} that a necessary and sufficient condition for having fast instabilities is an angular crossing in $g_{E,\mathbf{v}}$ defined in Eq. \ref{eq:gp}. Assuming to have azimuthal symmetry around a specific direction, angular crossing means that there is a  $v_0$ for which $g_{E,v>v_0} \times g_{E,v<v_0}<0$. In  \cite{Dasgupta:2021gfs} such a proof has been made more general: flavor instabilities, either fast or slow, can arise only if there is a spectral crossing, either in angle or energy.

In the following we review the state of the art for both slow and fast instabilities. Other reviews focused on slow modes are \cite{Mirizzi:2015eza,Chakraborty:2016yeg}, whereas the only one already available on fast modes is \cite{Tamborra:2020cul}.

\subsubsection{Slow Flavor Conversions}
\label{sec:Slow_Conversions}

The seminal papers investigating collective effects assumed the so called bulb model, where neutrinos are emitted uniformly and half-isotropically from the surface of a spherical neutrinosphere. Moreover, neutrino emission is taken to be azimuthally symmetric and that the conditions in the star depend only on the radial distance. In this framework, Eq. \ref{eq:boltz} becomes a system of first order ordinary differential equations that has been solved numerically by different groups \cite{Duan:2005cp,Duan:2006an,Duan:2006jv,Hannestad:2006nj,Fogli:2007bk,Duan:2007mv}. Some understanding can already be obtained analytically  through  an  analogy  with  a gyroscopic  pendulum  in  flavor  space \cite{Hannestad:2006nj,Fogli:2007bk,Duan:2007mv}.  In normal mass ordering  the  pendulum  starts  in a downward (stable) position where the small value of the mixing angle induces only negligible  flavor  changes.   Conversely,  in inverted mass ordering the  pendulum  starts  in  upward  (unstable)  position in and  it induces  maximal  flavor conversions $\nu_e\bar{\nu}_e\to\nu_x\bar{\nu}_x$,  conserving lepton  number. The main feature in this context is a spectral split: $\bar{\nu}_e$ are completely converted to $\bar{\nu}_x$, whereas for $\nu_e$ this happens for $E>E_c$. This happens because the the initial energy spectra of $\nu_e$ are lager than the ones of $\bar{\nu}_e$ and $\bar{\nu}_x$ and the total lepton number must be conserved.

For a while the presence of spectral splits \cite{Duan:2007bt,Raffelt:2007cb,Fogli:2008pt,Raffelt:2007xt,Galais:2011gh} in inverted ordering and the absence of significant flavor conversions in normal ordering has been considered to be the paradigm for slow conversions. It was later realized that the outcome strongly depends on the many assumptions entering the calculations. Here we briefly provide a list of them.

\begin{itemize}
\item \textbf{Flavor asymmetries \cite{Esteban-Pretel:2007jwl,Fogli:2009rd,Chakraborty:2014lsa,Dasgupta:2009mg}}. The absence of conversions in normal ordering seems to hold only for a large hierarchy between neutrino fluxes ($f_{\nu_e}\gg f_{\bar{\nu}_e}\gg f_{\nu_x}$, typical of the accretion phase), whereas for more similar fluxes ($f_{\nu_e}\gtrsim f_{\bar{\nu}_e}\gtrsim f_{\nu_x}$) one can find multiple splits and also in normal ordering.

\item \textbf{Large matter potential \cite{Chakraborty:2011nf,Chakraborty:2011gd,Dasgupta:2011jf,Sarikas:2011am,Saviano:2012yh,Chakraborty:2014nma}}. A matter potential larger than the neutrino-neutrino interaction potential leads to a suppression of flavor conversions. This is typically happening closer to the supernova core and during the accretion phase.

\item \textbf{Multi angle effects \cite{Mirizzi:2010uz,Cherry:2010yc,Mirizzi:2011tu,Mirizzi:2012wp}}. When the flavor asymmetries are mild the phase dispersion induced by different propagation lengths of neutrinos can smear or completely remove the effects of the spectral splits, eventually leading to complete decoherence, i.e. all flavors are equilibrated up to lepton number conservation.

\item \textbf{Three flavors effects \cite{Dasgupta:2007ws,Duan:2008za,Fogli:2008fj,Dasgupta:2010ae,Friedland:2010sc,Dasgupta:2010cd}}. As for the multi angle effects, small flavor asymmetries can induce complete spectral swaps or even flavor equilibration among all three flavors.

\item \textbf{Sponteneous breaking of azimuthal \cite{Raffelt:2013rqa,Raffelt:2013isa,Duan:2013kba,Mirizzi:2013rla,Chakraborty:2013wsu}, spatial homogeneity \cite{Mangano:2014zda,Duan:2014gfa,Mirizzi:2015fva,Mirizzi:2015hwa,Duan:2015cqa,Abbar:2015fwa,Chakraborty:2015tfa} and stationarity \cite{Dasgupta:2015iia,Capozzi:2016oyk}.} It has been realized that the symmetries used as initial conditions of neutrino emission (azimuthal symmetry, spatial homogeineity, stationarity) are spontaneously broken during neutrino propagation. This leads to new instabilities, that can develop in both mass orderings, but that are in general suppressed when the matter potential is dominating. Nevertheless, it has been shown \cite{Dasgupta:2015iia,Capozzi:2016oyk} that  self-interacting neutrinos can generate a pulsating solution with a frequency that effectively compensates the phase dispersion associated with the large matter term, lifting the suppression and making collective oscillations possible deep inside the supernova. However, because the matter potential changes during neutrino propagation, it is not clear whether a flavor wave with a specific pulsation can have enough time to grow and lead to significant flavor conversions. The presence of turbulent variations of the matter potential may introduce a coupling among flavor waves with different $k$, so making it more likely to have an instability even when neutrinos are propagating away in a supernova \cite{Abbar:2020ror}.

\item \textbf{Neutrino halo effect \cite{Cherry:2012zw,Cherry:2013mv,Sarikas:2012vb,Zaizen:2019ufj,Cherry:2019vkv,Zaizen:2020mun}}. Neutrinos  are  not completely free-streaming after the neutrinosphere and even a  small fraction of scattering neutrinos can produce a small ``neutrino halo". Such inward going neutrinos can modify the outcome of conversions and the shape of spectral splits, if present, but, according to the latest simulations, the effects have been found to be relatively small.

\end{itemize}

\subsubsection{Fast Flavor Conversions}
\label{sec:Fast_Conversions}

For slow conversions it is usually assumed that neutrinos are emitted half isotropically, making it intrinsically impossible to create an angular crossing in $g_\mathbf{p}$. The importance of a non-trivial angular distributions of neutrinos was first recognized in \cite{Sawyer:2005jk,Sawyer:2008zs,Sawyer:2015dsa}. It was realized that angular crossings can lead to new instabilities, developing on extremely short time scales, and that can occur even in the absence of neutrino mixing. In this case, flavor conversions is entirely due to pairwise interactions of the type $\nu_e(\mathbf{p})+\nu_x(\mathbf{q})\to\nu_e(\mathbf{q})+\nu_x(\mathbf{p})$ and $\nu_e(\mathbf{p})+\bar{\nu}_e(\mathbf{q})\to\nu_x(\mathbf{q})+\bar{\nu}_x(\mathbf{p})$. The potential impact that a change of flavors happening on such short time scales can have on the explosion and on the nucleosynthesis of heavy nuclei has lead to an extensive research activity by the entire community, which is still underway.

Some work has provided a detailed characterization of fast conversions using the linear stability analysis \cite{Izaguirre:2016gsx,Capozzi:2017gqd,Yi:2019hrp,Airen:2018nvp,Capozzi:2019lso,Chakraborty:2019wxe}. Useful insights has also been obtained analytically, usually working under the assumption of spatial homogeneity and azimuthal symmetry \cite{Dasgupta:2017oko,Johns:2019izj,Padilla-Gay:2021haz}. In particular in \cite{Padilla-Gay:2021haz} it has been pointed out that the analogy with a pendulum in flavor space works also in the case of fast conversions. However, differently  from what happens for slow conversions, the real part of the  pulsation of the flavor wave resulting from the linear analysis acts as the pendulum spin, and plays role in determining the final amount of conversions.

With a significant development of the stability analysis, it has become possible to apply this tool directly to data provided by supernova hydrodynamical simulations, with the purpose of assessing whether the conditions for fast flavor instabilities are actually satisfied. This is the closest one can get to knowing whether fast instabilities are possible in real supernovae. However, simulations usually do not provide the angular distributions of the neutrino fluxes, but only their angular moments defined as:
\begin{equation}
\mathbf{I}_n=\int_{-\infty}^{+\infty}\frac{E^2dE}{2\pi^2}\int \frac{d\mathbf{v}}{4\pi}\,g_{E,\mathbf{v}}\mathbf{v}^n\,.
\label{eq:angular_moments}
\end{equation}
A few methods \cite{Dasgupta:2018ulw,Abbar:2020fcl,Nagakura:2021txn,Nagakura:2021suv,Nagakura:2021hyb}, have been proposed to perform a stability analysis using only the angular moments $\mathbf{I}_n$ and they have been recently applied to a plethora of supernova simulations. The first investigations in \cite{Tamborra:2017ubu,DelfanAzari:2019epo} found no evidence of instabilities, but they were either limited to 1D simulations or to very specific time and space locations of the supernova. More recent studies reported that crossings are possible in the following areas of a supernova:

\begin{itemize}
\item \textbf{Proto-neutron star \cite{DelfanAzari:2019tez,Abbar:2019zoq,Glas:2019ijo}.} The physical origin of the crossings is a strong convective activity happening in the proto-neutron star, which can generate large amplitude modulations in the spatial distributions of $\nu_e$ and $\bar{\nu}_e$ number  densities.   The  physical  implications  are not very clear due to the nearly equal distributions of neutrinos and antineutrinos of all flavors.

\item \textbf{Neutrino decoupling region \cite{Abbar:2018shq,Nagakura:2019sig,Abbar:2019zoq}}.   Their  existence  can be explained by the  neutron  richness  of  matter, which induces a later  decoupling of $\nu_e$ with respect to $\bar{\nu}_e$ and, thus, a more forward peaked angular  distributions  of   $\bar{\nu}_e$. Another possibility  is the presence of LESA \cite{Tamborra:2014aua,OConnor:2018tuw,Glas:2018vcs,Nagakura:2019evv}, i.e. an asymmetric emission of lepton number, or other phenomena \cite{Nagakura:2019evv}.

\item \textbf{Free streaming regime \cite{Morinaga:2019wsv,Capozzi:2020syn,Abbar:2020qpi}.} Crossings can be generated by neutrino backward scatterings off heavy nuclei and their size seems to get larger for smaller radial distances  \cite{Morinaga:2019wsv,Capozzi:2020syn,Abbar:2020qpi}. Such crossings are ubiquitous in the pre-shock region, but they can also occur in the post-shock region.  In the former case, there is hardly an impact on astrophysical processes and on the detection at Earth \cite{Zaizen:2021wwl,Abbar:2021lmm}, both because of the slower growth rates of the instabilities and the very small size of the amount of conversions expected. On the other hand, the instabilities developing in the post-shock region might produce an observable effect.

\end{itemize}

Even if only under some simplifying assumptions, fast flavor instabilities have been also simulated numerically in order to assess what is the final amount of conversions expected and how it depends on the initial conditions. It was originally thought that such instabilities lead to equilibration among all flavors \cite{Sawyer:2005jk,Sawyer:2008zs,Sawyer:2015dsa,Dasgupta:2016dbv}, up to conservation of the total lepton number. This conclusion was considered to hold for all neutrino energies and mass orderings, but it has been recently called into questions. Here we report a brief summary of the findings obtained through numerical simulations:
\begin{itemize}
    \item \textbf{Neutrino propagation \cite{Shalgar:2019qwg}.} Even if a system of neutrinos has an initial crossing and thus develops a fast flavor instability, the propagation in space and time may cancel the crossing \cite{Shalgar:2019qwg}.
    
    \item \textbf{Propagation of the power of the instability to small angular scales \cite{Abbar:2018beu,Martin:2019gxb,Johns:2020qsk,Bhattacharyya:2020dhu,Bhattacharyya:2020jpj,Wu:2021uvt}.} During its evolution, the power of a fast instability is moved from large scales to small ones in momentum space, accelerating the decoherence of the system (and the equilibration of flavors) \cite{Abbar:2018beu,Martin:2019gxb,Johns:2020qsk,Bhattacharyya:2020dhu,Bhattacharyya:2020jpj,Wu:2021uvt}. Moreover, when considering a coarse-graining over space, fast conversions seem to reach eventually a steady state \cite{Bhattacharyya:2020dhu,Bhattacharyya:2020jpj,Wu:2021uvt}. In particular in \cite{Bhattacharyya:2020jpj} an approximate analytical formula has been derived for calculating the amount of decoherence reached in fast conversions and, thus, the final amount of flavor conversions. This formula depends on the propagation angle of neutrinos and on the initial asymmetry between the total number of neutrinos and antineutrinos. If the asymmetry is extremely small, as expected for the crossings generated by backward scattering of neutrinos in the free streaming regime, then the amount of flavor conversions is expected to be negligible, as confirmed in \cite{Zaizen:2021wwl,Abbar:2021lmm}. In \cite{Bhattacharyya:2020jpj} the analytical formula is independent from the type of perturbation used to seed the flavor instabilities. However, in \cite{Wu:2021uvt} a distinction is made between localized and random seeds. In the first case a coherent behaviour in the space and momentum evolution of the flavor wave is retained for a longer time. The difference between \cite{Bhattacharyya:2020jpj} and \cite{Wu:2021uvt} can be associated to heterogeneous numerical methods employed for calculating spatial derivatives.
    
    \item \textbf{Spontaneous symmetry breaking \cite{Shalgar:2021wlj}.} As it happens to slow conversions, also in the context of fast ones there is a spontaneous breaking of the symmetries imposed in the initial conditions. This has been shown to happen for the azimuthal symmetry in \cite{Shalgar:2021wlj} and in \cite{Capozzi:2020kge}, though in the second reference it was not explicitly stated in the conclusions.

    \item \textbf{Dependence on neutrino energy \cite{Shalgar:2020xns}.} It has been proposed that the outcome of fast conversions depend on the size of the mass differences between mass eigenstates and on their ordering \cite{Shalgar:2020xns}. This claim has been criticized in \cite{Martin:2021xyl} where only a modest dependence has been observed. However, the first simulation has been performed considering an homogeneous system, whereas the second has taken into account also the spatial evolution.
    
    \item \textbf{Impact of inelastic collisions \cite{Capozzi:2018clo,Shalgar:2020wcx,Sasaki:2021zld,Martin:2021xyl,Sigl:2021tmj}.} Since the conditions for fast conversions have been found even in locations where neutrinos are still partially or completely coupled to the plasma, there have been a few studies implementing collisions in numerical simulations. In this context  \cite{Shalgar:2020wcx,Sasaki:2021zld} reported the possibility of enhancement of flavor conversions, assuming only evolution in time. On the other hand in \cite{Martin:2021xyl,Sigl:2021tmj} both time and space evolution has been taken into account and it was observed that collisions might cause flavor depolarization, thus suppressing conversions, but that a much larger mean free path than expected is required in order for this happen. This is in agreement to what found in \cite{Capozzi:2018clo}, where the role of collisions in the generation of crossing has been also pointed out. 
    
    \item \textbf{Dependence on the number of crossings \cite{Bhattacharyya:2021klg}.} Multiple crossings can inhibit flavor conversions \cite{Bhattacharyya:2021klg}. This is especially relevant when considering fast conversions in the early universe, where the neutrino angular distributions are expected to be more similar compared to the supernova case.
    
    \item \textbf{Dependence on the number of neutrino species \cite{Capozzi:2020kge,Shalgar:2021wlj}.} Considering six neutrino species the crossings and, thus, fast instabilities can occur in one (or more) of the three sectors $e\mu,e\tau,\mu\tau$ and then propagate to other ones \cite{Capozzi:2020kge}. Moreover, even considering the distributions of $\nu_\mu,\nu_\tau,\bar{\nu}_\mu,\bar{\nu}_\tau$ to be the same, the outcome of flavor conversions is different \cite{Shalgar:2021wlj} with respect to what obtained by using only the equation of motions for three species, as usually done in literature.
    
    \item \textbf{Impact of new physics \cite{Dighe:2017sur}.} Fast conversions can be affected by the presence of new physics, such as non-standard interactions \cite{Dighe:2017sur}.

\end{itemize}

\section{Fast Flavor Conversions in Compact Binary Mergers}

\label{sec:Compact_binaries}

Neutrinos are copiously produced also in merger events between two neutron stars or a neutron star and a black hole, reaching a total luminosity of $\sim10^{53}$ erg \cite{Ruffert:1996by,Foucart:2015vpa}. Despite their low chance of detectability, these neutrinos play a fundamental role in cooling the accretion torus that is created around the central remnant (a massive neutron star or a black hole), and the nucleosynthesis of heavy nuclei through $r$-processes. \cite{Ruffert:1996by,Foucart:2015vpa,Kulkarni:2005jw,2010MNRAS.406.2650M,Perego:2014fma}. Therefore, as it happens for core-collapse supernovae, a full understanding of neutrino flavor conversions is mandatory.

A peculiarity of such environments is the excess of $\bar{\nu}_e$ over $\nu_e$. A first consequence is the possible occurrence of  the ``matter-neutrino resonance" \cite{Malkus:2014iqa,Malkus:2015mda,Wu:2015fga}, i.e. the almost complete cancellation of the neutrino self interaction potential $\Omega_\mathbf{p}^{\nu\nu}$ with the potential due to interactions with electrons $\Omega_\mathbf{p}^{\rm mat}$. A second consequence is the ubiquity of crossings and thus of fast flavor instabilities in the accretion disk, first recognized in \cite{Wu:2017qpc}. Indeed, as it happens in supernovae, $\nu_e$ are coupled to the plasma in a larger volume, thus, despite the overall excess of $\bar{\nu}_e$, for specific backward going propagation directions there may be a local excess of $\nu_e$. However, the large region of instability reduces its size after $O(10)$ ms from the formation of the accretion disk. This time interval can be larger in case of a massive neutron star remnant \cite{George:2020veu}. Following this finding, the impact of fast instabilities on nucleosynthesis has been assessed in \cite{Wu:2017drk} and \cite{George:2020veu}. Both references assumed complete flavor equilibration. The first found that the fraction of lanthanides produced in the neutrino-driven wind may be increased by a factor of $O(10^3)$. On the other hand, the second reference found an enhancement of the  iron  peak  abundances  and  a reduction of  the  first  peak  abundances,  whereas lanthanides are marginally affected. Flavor equilibration was assumed also in \cite{Li:2021vqj}, but, differently from the previous works, a magneto-hydrodynamic simulation of a typical post-merger accretion disk was used. Their results show that fast conversions can  increase the production of  the  lanthanides  and of elements in the third peak to a level comparable  to  solar  abundances.   This  is an interesting  hint for  massive  post-merger  disks  being  a  major production site for heavy elements. 

All the results on the impact of fast conversions on nucleosynthesis in post-merger accretion disks are based on the assumption of flavor equilibration. However, as discussed in the context of supernova neutrinos, such assumption might not always be justified. For instance, in \cite{Padilla-Gay:2020uxa} a simplified toy model was studied numerically, finding a conversion of less than 1\%. despite the large growth rate of fast instabilities.  

\section{Conclusions and future perspectives}
\label{sec:conclusion}

Neutrino flavor conversions are relevant in cosmological and  astrophysical dense environments, such as  the early universe, core-collapse supernovae and the merger of compact objects. Indeed, a full understanding of these phenomenona is mandatory for a correct interpretation of both the corresponding neutrino signals and (or) the astrophysical processes developing in these environments. The flavor content of a system of neutrinos is usually represented by a density matrix $\varrho_{\mathbf{p,x}}$, whose time and space evolution are described by Eq. \ref{eq:boltz}, where $\Omega_\mathbf{p,x}^{\nu\nu}$ represents the potential stemming from neutrino interactions among themselves. Such a term makes the evolution equation a non linear system of coupled partial differential equations, which has never been solved in its entire form. 

The early universe represents  a peculiar environment for testing nonlinear neutrino oscillations in high-density conditions. Indeed, the primordial nucleosynthesis (BBN), the leptogenesis, the cosmic background radiation (CMB), the formation of cosmological structures (LLS) can be profoundly influenced by the presence of neutrinos, by their properties such oscillations, interactions and non-zero mass. A special and interesting case is represented by the oscillations of an between active neutrinos and a sterile one, where the latter has a mass at the eV scale. The presence of such a light new degree of freedom is suggested by intriguing, but controversial, indications from short-baseline neutrino oscillation experiments. These sterile neutrinos are produced through mixing and contribute to the radiation content beyond photons and ordinary neutrinos. They also leave an imprint on different cosmological observables. In order to determine their relic abundance an accurate solution of Eq. \ref{eq:boltz} is mandatory.  For the mass and mixing parameters suggested by experimental analysis, the eV sterile neutrinos should be produced with the same number density of the active states. This result is however in disagreement with recent cosmological data on the radiation and on neutrino mass constraints. Extra ingredient should be added in the model of the flavor evolution in order to try to alleviate this tension. A full multi-flavor and multi momentum treatment of the active-sterile flavor evolution is mandatory in order to confirm or reject the eV sterile neutrino hypothesis. At the moment, the eV scenario is strongly disfavored by cosmological observations.

Sterile neutrinos with mass of the order of keV can represent natural dark matter candidates. 
A complete characterization of the production mechanisms for keV sterile neutrinos is extremely complex requiring the inclusion of hadronic contributions still not perfectly known, detailed solutions of the kinetic equations as well as some model building.  The results obtained should be
then compared with astrophysical and cosmological observables. This comparison
requires the use of signatures from different physical probes (X-ray, counting galaxy, Lyman-alpha) and
different distance scales. New  cosmological and astrophysical data will help shed light on the keV sterile neutrino paradigm.

Differently from the early universe, in the context of core-collapse supernovae and mergers of compact objects, flavor conversions are not fully understood even in the context of the Standard Model. In these cases different types of flavor instabilities can develop according to the shape of $g_{E,\mathbf{v}}$ (Eq. \ref{eq:gp}), i.e. the difference of the energy and angular distributions of neutrinos and antineutrinos, which are usually not equal to each other. Two approaches are usually employed to study these instabilities: linearized analyses and numerical solutions of a simplified version of Eq. \ref{eq:boltz}. The first tool is  computationally more manageable, but it only allows to know whether a flavor instability can develop given a set of initial conditions for $\varrho_{\mathbf{p,x}}$. Its application has shown that instabilities requires that $g_{E,\mathbf{v}}$ has a crossing either in energy $E$ or in the propagation direction $\mathbf{v}$. Instabilities can be classified according to their growth rate: slow conversions grow with a rate $\propto \sqrt{n_\nu \omega_E}$, whereas fast conversions with $\propto n_\nu$. The latter can develop on time scales as small as few nanoseconds, if they occur close to the neutrinosphere, with a potential major impact on the supernova explosion. The application of the linear stability analysis on real data provided by hydrodynamical simulations has shown that crossings are indeed a relatively common feature in supernovae. These searches have been performed by several groups on different data and a general consensus has been achieved. Nevertheless, a systematic study of how crossing generation depends on the characteristics of supernova progenitors is still lacking, especially for three-dimensional models which are in principle the most realistic representation of the real explosion event, but at the same time are more time consuming in terms of simulations.

A lot of work has been dedicated to numerically  solving Eq. \ref{eq:boltz} and it is now clear that the results obtained can deviate from the real non linear dynamics of the system when the equation is simplified by imposing some symmetries. This has been observed for slow conversions, for which the spectral splits originally found within the bulb model seem to be washed out when allowing space or time instabilities to develop.  Similarly, it was thought that fast conversions lead to flavor equilibration, but more recently the dependence of this result on the presence of collisions, the symmetries of the system and the neutrino energy has been highlighted. Nevertheless, results obtained by different calculations are not always in perfect agreement. Future studies should continue investigating the phenomenology of flavor conversion when symmetry assumptions are relaxed, as well as the possible interplay between fast and slow conversion (and non standard physics), which is still largely unknown. Ultimately, an implementation of flavor conversions into hydrodynamical simulations is necessary in order to assess what is the impact on the astrophysical processes happening in the exploding star. This is of course computationally extremely challenging, if not even unfeasible at the moment. However, a first step in this direction can be similar to the one adopted for mergers of compact objects, where complete flavor conversions has been assumed at those locations in space and time where conditions for flavor instabilities are satisfied. A more realistic approach can be either moving away from flavor equilibration, or writing Eq. \ref{eq:boltz} in terms of the angular and energy moments of the density matrix $\varrho_{\mathbf{p,x}}$. The latter possibility is numerically more manageable and it has been recently adopted \cite{Myers:2021hnp}. Finally, we underline the importance of cross checks between different numerical codes.

\section*{Acknowledgements}
We thank Alessandro Mirizzi for valuable suggestions.
The work of FC is supported by GVA Grant No.CDEIGENT/2020/003.
This work was partially supported by the research grant number 2017W4HA7S “NAT-NET: Neutrino and As- troparticle Theory Network” under the program PRIN 2017 funded by the Italian Ministero dell’Universit`a e della Ricerca (MUR). The authors also acknowledge the support by the research project TAsP (Theoretical Astroparticle Physics) funded by the Istituto Nazionale di Fisica Nucleare (INFN).

\bibliographystyle{bibi}
\bibliography{Bibliography}

\end{document}